\documentstyle[aps,eqsecnum,preprint,tighten,floats]{revtex}

\def\be{\begin{equation}}
\def\ee{\end{equation}}

\begin{document}
\draft

%
%
\title{ELECTRO-MAGNETIC NUCLEON FORM FACTORS\\
AND THEIR SPECTRAL FUNCTIONS IN SOLITON MODELS}
\author{G.~Holzwarth\thanks{%
e-mail: holzwarth@hrz.uni-siegen.d400.de}
}
\address{Fachbereich Physik, Universit\"{a}t-GH-Siegen, 
D-57068 Siegen, Germany} 
%
%
\maketitle
\begin{abstract}
It is demonstrated that in simple soliton models 
essential features of the electro-magnetic nucleon form factors 
observed over three orders 
of magnitude in momentum transfer $t$ are naturally reproduced.  
The analysis shows that three basic ingredients are required: 
an extended object, partial coupling to vector mesons, and 
relativistic recoil corrections. We use for the extended object
the standard skyrmion, one vector meson propagator
for both isospin channels, and the relativistic boost to the Breit
frame.
Continuation to timelike $t$ leads to quite stable results for the
spectral functions in the regime from the 2- or 3-pion threshold to
about two rho masses. Especially the onset of the continuous part
of the spectral functions at threshold can be reliably determined
and there are strong analogies to the results imposed on dispersion
theoretic approaches by the unitarity constraint.
\end{abstract}
%

\newpage


\section{Introduction}

Topological soliton models for structure and dynamics of baryons
are based on effective nonlinear lagrangians for selected 
mesonic degrees of freedom. These usually comprise the  
pseudoscalar Goldstone boson octet of spontaneously broken chiral symmetry,
but also the light vector and axial vector mesons have been
included.

A decisive advantage of the soliton concept as compared to models
where explicit pointlike fermion fields are coupled to meson and gauge fields
is the fact that already in leading classical approximation
the spatial structure of the baryon as an extended object
is obtained from the underlying effective action. Therefore all types of
form factors can readily be extracted from the models, and comparison
with the experimentally observed dependence on momentum transfer
presents a stringent test for the resulting spatial profiles.
Specifically, the electro-magnetic form factors of the nucleon
for which we expect a wealth of precise data in the few (GeV/c) region
from the new generation of electron accelerators pose a severe
challenge for chiral soliton models. 

Shortly after the initial work of~\cite{Braaten} the e.m. nucleon
form factors have been evaluated for various versions of effective
meson models~\cite{MKW87,KVWM88,Mei93} and various sets of parameters, 
mainly for
momentum transfers $Q<$ 1 GeV/c. The conclusion was that the resulting
$Q^2$-dependence follows roughly the standard dipole form, and that
reasonable values for radii require the presence of explicit vector
mesons. Unfortunately, in cases where these results have been included in 
more detailed comparisons with new data (e.g. in~\cite{Pfeil,Locher}), they
have left the prejudice that soliton form factors are not satisfactory. 
However, for momentum transfers $Q>$ 1 GeV/c it is important to incorporate 
relativistic kinematical corrections into the form factors. This is
crucial for quark bag or cluster 
models~\cite{Licht,Foster,Barnhill,boobag,Pfeil,Locher} 
as well as for soliton
models~\cite{Ji91}. The implementation of these corrections
is fairly easy for solitonic nucleons
due to the Lorentz covariance of the underlying field equations.

The proton magnetic form factor is quite accurately known up to 30
(GeV/c)$^2$ and it is not difficult to show~\cite{Eri} 
that with suitable choice of 
coupling constants the relativistically corrected soliton form factor 
calculated from a minimal vector meson model
follows the data closely over more than three orders of
magnitude in $Q^2$. This may be a bit disappointing because one might
have hoped that in this region of high momentum transfer the 
characteristic shape of form factors could reflect dynamical
features of QCD~\cite{QCDcorr}. On the other hand, 
a similar conclusion had already
been reached by H\"{o}hler~\cite{Hoe76} through a dispersion-theoretic
fit of the spectral functions which included a continuous part to
account for the Frazer-Fulco unitarity constraint~\cite{FraFu}.

In the following we would like to analyze which basic qualitative features
of the e.m. form factors and their spectral
functions can be expected from the soliton concept for nucleons and to
which extend these may correspond to observed facts. 
In order to expose the typical features of solitonic e.m. form factors
in the most transparent way it is necessary to choose the
essential ingredients as simple as possible. We therefore do not strive
for optimal agreement with the data for some sophisticated model, but instead
choose the most simple construction:
$i$) a purely pionic soliton created through the standard Skyrme term;
this represents the extended object with its spatial structure defined
through the chiral profile $F(r)$; $ii$) its minimal coupling to the 
electro-magnetic field is partly mediated through vector mesons of 
mass $m_V$ represented by one Klein-Gordon propagator; 
$iii$) relativistic boost factors to the Breit frame account for kinematic
corrections. 

In section II we show that these three ingredients are sufficient to
produce for $Q^2<$ 1 (GeV/c)$^2$ perfect agreement with the standard dipole
and for $Q^2>$ 1 (GeV/c)$^2$ the characteristic deviations from the
dipole form. We also get indications at which point more precise data 
could lead to discrepancies. 

Of course, it is most interesting to try to learn something about the
spectral functions which underlie the resulting form factors. Especially
intriguing is the question if and what the Skyrme soliton knows about
the Frazer-Fulco constraint. A direct investigation of this problem
by relating the pion-baryon phase shifts to the timelike form factor
appears difficult, because the $\pi$-$N$ S-matrix in the soliton 
formulation depends only on one variable. 
Earlier analyses of the form factors for
timelike momentum transfers~\cite{KVWM88} are not helpful in this respect
because the method of analytic continuation used there~\cite{Bowcock}, 
provides only 
averaged form factors with an averaging interval that conceals all
detailed structures. 
Unfortunately, even within this simple model the
continuation into the timelike regime remains ill-posed; but it seems
that obtaining the spectral functions as Laplace inverse of the spatial
densities provides rather stable results in the most interesting region
from threshold to about 2 $m_\varrho$. We discuss this procedure in
detail in section III for the electric form factors; the magnetic
spectral functions then are easily obtained by differentiation or
integration. Once the spectral functions of the plain skyrmion are
determined in its restframe, the modifications due to
partial coupling to vector mesons, and the boost to the Breit frame
pose no additional problem. They are discussed in sections IV and V.

\section{Relativistic form factors in the soliton model}

We create the nucleon as topological soliton
through the standard Skyrme model~\cite{Sk61} ($U\in$ SU(2))
\be
{\cal L}(U)={\cal L}^{(2)}+{\cal L}^{(4)}
\label{eq:1} \ee
\be
{\cal L}^{(2)}=\frac{f_\pi^2}{4}\int \left(-trL_\mu L^\mu+m_\pi^2 
tr(U+U^\dagger-2) \right)d^3x.
\label{eq:2} \ee
\be
{\cal L}^{(4)}=\frac{1}{32e^2}\int tr[L_\mu,L_\nu]^2 d^3x
\label{eq:3}  \ee
where $L_\mu$ denotes the chiral gradients 
\be
L_\mu = U^\dagger \partial_\mu U .
\label{eq:4}  \ee
The pion decay constant $f_\pi$ and the pion mass $m_\pi$ have their 
physical values $f_\pi$=93 MeV and $m_\pi$=138 MeV, the Skyrme
parameter $e$ is taken at $e$=4.25 which leads to the 
correct Delta-nucleon mass difference of $M_\Delta-M_N$=295 MeV.

The static soliton solution, the Skyrme hedgehog $U=\exp(i\vec\tau\cdot\hat
rF(r))$ is characterized by the numerically determined chiral profile
$F(r)$ which carries all the information about the spatial structure
of the extended object.

In the Breit frame for spacelike $t=-Q^2$ the isoscalar ($T$=0) and
isovector ($T$=1) electric and magnetic (Sachs) form
factors are given as Fourier transforms of baryon density $B_0(r)$ (for
$T$=0) and moment-of-inertia density $B_1(r)$ (for $T$=1): 
\begin{eqnarray}
 G^{T=0}_E (-Q^2) & = & \frac{1}{2}\int d^3r\: j_0(Qr) B_0(r)
\label{eq:5} \\
 G^{T=0}_M (-Q^2) & = & \frac{3}{2\:r_B^2} \int d^3r \: \frac{j_1(Q
r)}{Qr}\: r^2 B_0(r)
\label{eq:6} \\
 G^{T=1}_E (-Q^2) & = & \frac{1}{2} \int d^3r\: j_0(Qr) B_1(r)
\label{eq:7} \\
 G^{T=1}_M (-Q^2) & = & \frac{3}{2} \int d^3r\: \frac{j_1(Qr)}{Qr} B_1(r)
\label{eq:8}
\end{eqnarray}
Expressed in terms of the chiral profile $F(r)$ 
the densities $B_0(r)$ and $B_1(r)$ are 
\be
B_0(r) = - \frac{1}{2\pi^2} F'\; \frac{\sin^2F}{r^2},
\label{eq:9}  \ee
\be
B_1(r) = \frac{2}{3\Theta} \left[f^2_\pi \sin^2 F
+ \frac{1}{e^2} \sin^2 F \left( F^{'2} +
\frac{\sin^2F}{r^2} \right) \right]
\label{eq:10}  \ee
with normalization
$\int d^3r \;B_0(r)$ =1, and $\int d^3r \;B_1(r)$ =1.

For the magnetic form factors the isoscalar and isovector magnetic moments
\be
2\mu_0=\mu_p+\mu_n=\frac{M_N}{3\Theta}r_B^2, \mbox{~~~~~~~~~~}
2\mu_1=\mu_p-\mu_n=\frac{2}{3}M_N \Theta
\label{eq:11}  \ee
(with baryonic square radius $r_B^2=\int d^3r\; r^2 B_0(r)$)
have, respectively, been divided out such that they both satisfy the
same normalization condition as their electric counterparts
\be
G_M^0(0)=G_M^1(0)=\frac{1}{2}.
\label{eq:12}  \ee

The Skyrme term ${\cal L}^{(4)}$ in (\ref{eq:3}) effectively represents the
lowest local approximation to explicit inclusion of higher
resonances, notably $\varrho$ and $\omega$ 
mesons~\cite{JJMPS88,MKWS89,SWHH89}. Propagating
vector mesons introduce additional pole structures near the vector
meson mass into the form factors, which cannot be accounted for
by the local Skyrme term alone in a satisfactory way. However, instead of
dealing with the ambiguities and numerous parameters of vector meson
models it appears more appropriate to incorporate their effect into
one common factor $\Lambda(t)$ to be multiplied with the pure
Skyrme model form factors (\ref{eq:5})-(\ref{eq:8}),
\be 
\Lambda(t)=\lambda \left(\frac{m^2_V}{m^2_V-t}\right) +(1 -\lambda).
\label{eq:13} \ee
With the physical vector meson mass $m_V$=770 MeV this factor contains
one additional parameter $\lambda$ which allows to interpolate between
complete vector meson dominance ($\lambda$=1) and the purely pionic Skyrme
model ($\lambda$=0). The resulting form factors 
\be 
G_{E,M}^{0,1}(t) = \Lambda(t) \;
G_{E,M}^{0,1(\pi)}(t) 
\label{eq:14} \ee
can be shown to represent good approximations to results
obtained in models with explicit vector mesons included. (For the
coupling to the $\omega$-meson in the minimal model it is exact.)

Although the definitions (\ref{eq:5})-(\ref{eq:8}) hold 
in the Breit frame only, expressions (\ref{eq:9}) 
and (\ref{eq:10}) for the densities contain the chiral profile
$F(r)$ as obtained in the soliton restframe. However, 
both frames coincide only for infinite soliton mass $M_S$. 
For values of $Q^2$ near and beyond the actual
soliton mass $ M_S $ kinematic corrections have to be included 
by employing the hedgehog soliton boosted to the Breit frame, where it moves
with velocity $v$
\be
\gamma^2 = (1-v^2)^{-1} = 1 + \frac{Q^2}{4 M^2} = 1 -
\frac{t}{4M^2} 
\label{eq:15} \ee
with spacelike momentum transfer $ t= - Q^2 \leq 0$. 

An approximate way to implement these recoil corrections has been 
suggested by Ji~\cite{Ji91}. The electric form factors are Lorentz scalars, 
therefore one has
\be
G_E(t) = G^{(nr)}_E (t/\gamma^2)
\label{eq:16} \ee
where $G^{(nr)}_E$ is the non-relativistic form factor 
evaluated in the soliton restframe.

For the magnetic form factors the boost to the Breit frame introduces
an additional factor $\gamma^{-2}$
\be
G_M(t) = \frac{1}{\gamma^2} G_M^{(nr)} (t/\gamma^2) \; .
\label{eq:17} \ee
These boost factors are well-known in models for relativistic quark
clusters~\cite{Licht}. There, however, $G_E$ carries the same factor
$\gamma^{-2}$ as $G_M$ due to the normalization of the spectator quarks.

The kinematical transformations (\ref{eq:16}) and (\ref{eq:17}) 
determine the asymptotic behavior
of the relativistic form factors for large spacelike $t\to -\infty$:
\be
G_E(t) \to\;\;\left[ G_E^{(nr)}(-4M^2)\;+\;
\frac{4M^2}{\gamma^2}\frac{\partial}{\partial t}
G_E^{(nr)}(-4M^2)\right] \;+\; {\cal O}\left(\frac{1}{t^2}\right)
\label{eq:18} 
\ee
\be
G_M(t) \to\;\;\left[\frac{1}{\gamma^2} G_M^{(nr)}(-4M^2)\right]
\;+\; {\cal O}\left(\frac{1}{t^2}\right).
\label{eq:19} 
\ee
These asymptotic forms show a
rather undesirable feature of the boost transformations (\ref{eq:16}),
(\ref{eq:17}) because
they violate the superconvergence rule expected 
for e.m. form factors~\cite{LaBoe,countrule}
\be
\lim _{t \to -\infty} \; t G_{E,M}(t) =0.
\label{eq:20} \ee   
Practically, however, the absolute values of $G_{E,M}^{(nr)}$ 
and its derivative at $t=-4M^2$ are so small that up to $Q^2$=10
(GeV/c)$^2$ the influence of the undesired terms is not important.
This is demonstrated in figs.1,2
where the dashed lines show the e.m. form factors {\it after}
subtraction of the terms written out in square brackets in
eqs.(\ref{eq:18}), (\ref{eq:19}).

In principle, within tree approximation, the kinematic mass $M$ must be
identified with the classical soliton mass $M_S$. Ideally, of course,
$M$ should coincide with the physical nucleon mass $M_N$. However, it is
known, that this difference is related to quantum 
corrections~\cite{Mouss,MeiWa} which 
probably also affect the existence and positions of zeros in 
the non relativistic form factors.
Therefore, $M$ is not really well defined in the model and could be used
as additional parameter to minimize the undesired terms in
(\ref{eq:18}), (\ref{eq:19}) by
choosing $M$ such that $G_{E,M}^{(nr)}(-4M^2)=0$. 
In this way superconvergence could be enforced 
at least for $G_M$. However, to keep the number of parameters as
small as possible we strictly adhere to the tree approximation and
identify in the following the kinematical mass $M$ in the boost
transformation with the soliton mass $M_S$ which is $M_S$=1648 MeV
for the Skyrme model with $e$=4.25.   

Thus the only parameter remaining in the model is $\lambda$. It enters
sensitively into the radii and interpolates between the purely pionic
radii (which are generally too small) and the completely vector meson
dominated radii (which are too big):
\be
<r^2>=\frac{6}{G(0)}\:\frac{\partial}{\partial t}G(0) 
= \lambda \frac{6}{m_V^2}\;+\;<r^2>^{pionic}\;.
\label{eq:21} \ee
(For the electric neutron radius we take
\be
<r^2>_E^n=6\:\frac{\partial}{\partial t}G_E^n(0)=\frac{1}{2}(<r^2>_E^0
-<r^2>_E^1)\;; 
\label{eq:22} \ee
here the vector meson contributions cancel). 

The form factors shown in figs.1,2 are calculated with
$\lambda$=0.75, i.e. the e.m. coupling is strongly but not completely vector
meson dominated. The resulting square radii are (in fm$^2$):
$$
r_B^2=0.233;~~~~<r^2>_E^p=0.756;~~~~<r^2>_M^p=0.717;
~~~~<r^2>_E^n=-0.228;~~~~<r^2>_M^n=0.746; 
$$

Up to $Q^2\sim$ 0.5 (GeV/c)$^2$, $G_M^p$, $G_E^p$, $G_M^n$ follow the
standard dipole $G_D(t)=(1-t/0.71)^{-2}$ very closely. The magnetic 
form factors rise above the dipole in the region between 0.5 and 5   
(GeV/c)$^2$, and fall below the dipole above 5 (GeV/c)$^2$. The neutron
form factor $G_M^n$ rises slightly above the proton $G_M^p$, 
which seems to be in conflict with the data of~\cite{Lung} for $G_M^n$. 
In fig.2 for $G_M^n$ in addition to more recent data 
points~\cite{Lung,Marko,Bruins,Anklin} we have also included
a collection of older
data~\cite{akerlof,albrecht,rock,stein,hughes,dunning}. They seem to
indicate significant deviations from the standard dipole
parametrization also for small $Q^2$.  
But we find that at least up to 1 (GeV/c)$^2$ the ratio $G_M^p/G_M^n$
stays constant with good accuracy, and it appears difficult to
accommodate such differences between proton and neutron data.
With explicit dynamical inclusion of vector mesons it is not difficult to
find a parameter choice such that the rise of $G_M^p$ above the dipole 
coincides closely with that observed in the data (see e.g.~\cite{Eri}).
but it should be remembered that both, isoscalar and 
isovector magnetic moments $\mu_0$ and $\mu_1$
are known to be subject to sizable quantum corrections~\cite{MeiWa},
which may also affect the shape of the form factors.

In $G_E^p$ the corresponding rise above the dipole is suppressed 
and the form factor is dominated by a zero near 10 (GeV/c)$^2$. 
From our discussion of the boost transformation it is clear that the
high-$Q^2$ behavior of the ratios $G/G_D$ is sensitive to the position
of the first zeros in the form factors (relative to $4M^2$). From
(\ref{eq:5})-(\ref{eq:8}) 
we expect that the first zero in $G_E$ occurs at lower $Q^2$
than the first zero in $G_M$. (It may be noted that in the plain
Skyrme model $G_M$ is monotonously decreasing and has no zeros; 
however, inclusion of sixth-order terms or
explicit vector mesons leads to small fluctuations and corresponding 
zeros also in $G_M$). As long as these conditions are not 
inverted by quantum corrections we would therefore conclude
that the rapid fall-off in $G_E$ relative to $G_M$ is a typical and
rather stable feature of soliton models. A similar behavior is,
however, also observed in relativistic quark bag or cluster 
models (see e.g.~\cite{Barnhill,Pfeil,Locher}).  

Data points for the electric neutron form factor 
extracted from $e$-$d$ scattering
depend sensitively on the choice of the deuteron wave functions. 
This is illustrated in fig.2 for $G_E^n$ by the dotted
lines which represent fits of~\cite{Platchkov} of the form 
\be
G_E^n(-Q^2)=-a \mu_n \frac{Q^2}{4M_N^2}G_D(-Q^2)(1+b\frac{Q^2}{4M_N^2})^{-1}
\label{eq:23} \ee
where a and b depend on the choice of the n-p potential used in the
deuteron wave function.
Included in fig.2 is the data set  
of~\cite{Platchkov} for the Paris potential and an older data
collection of~\cite{Galster} based on the Feshbach-Lomon wave
functions. 
The dash-dotted line is the Galster fit~\cite{Galster} with
$a$=1, $b$=5.6. The shape of the calculated $G_E^n$ (full line) 
follows closely the Galster parametrization. 
(The absolute values of $G_E^n$ and thus the value of the slope near $Q^2=0$
can easily be reduced by allowing $\lambda$ to be
slightly different for isoscalar and isovector mesons. Again, however, 
without quantum corrections it is not very meaningful to adjust
parameters to the very sensitive electric neutron square radius, which 
experimentally is near $<r^2>_E^n$=$-0.12$ fm$^2$).
 
So, apart from choosing $\lambda$=0.75 we have made no 
attempt to further improve agreement with the experimental data.   
The results sufficiently demonstrate that this extremely simple model 
is capable to cover the essential features of the observed form
factors. Significant quantum corrections are expected to the magnetic moments,
to the radii, and to $M$. By the choice of $\lambda$ we have compensated
their possible influence on the radii. The high-$Q^2$ behavior of 
the form factors is sensitive to the choice of $M$ and
expected differences in loop corrections could
possibly be absorbed into slightly different values of $M$ for
different form factors. But here all form factors in figs.1 and 2
have been calculated with the same value of $M$ (the soliton mass).
As long as rotational and loop corrections to the currents
are not included, the accuracy shown in figs.1,2 is satisfactory and we
now turn to the typical features of the spectral functions which underlie
these form factors.

\section{Spectral functions for nucleon form factors in the 
nonrelativistic Skyrme model}

It is generally assumed that e.m. form factors $G(t)$ fulfill unsubtracted 
dispersion relations
\be
G(t) = \frac{1}{\pi} \int^\infty_{t_0} \frac{\Gamma (t')}{t' - t} dt'
\label{eq:24} \ee
where the spectral function $\Gamma(t)=$ Im $G(t)$ is given by the 
boundary value of $G(t+i0)$ along the upper edge of the cut which
extends from $t_0$ along the positive real axis in the $t$-plane.
The lower limit is $t_0=4m_\pi^2$ for isovector and $t_0=9m_\pi^2$ 
for isoscalar form factors. 
The determination of spectral functions from experimental form factor data 
in the spacelike region $t<0$ requires an analytical continuation
to the cut which is known to be mathematically an ill-posed
problem. Therefore such extrapolations have to be performed under
certain stabilizing assumptions and the resulting spectral functions
reflect these assumptions.

It is therefore of interest to find out what successful nucleon models 
may tell us about the structure of the spectral functions.
But it should be noted that (unless the models provide  analytical
expressions for the form factors) in practice we face the same ill-posed 
problem of inverting (\ref{eq:24}) with $G(t)$ given with finite {\it numerical} 
accuracy for all values $t<t_0$. However, apart from the fact that 
the numerical errors of the 'model data' for $G(t)$ can be made 
(arbitrarily) small, the models in general will
provide at least some analytical constraints on form factors (or on 
the underlying spatial densities) which will help to stabilize the 
analytic continuation.

In the following we analyze this procedure for the  
Sachs form factors of the simple Skyrme model.
Inclusion of additional explicit vector meson poles and
relativistic corrections then will be straightforward.

\subsection{Isoscalar electric $G_E^0(t)$}

We repeat that in the Breit frame for spacelike $t=-Q^2$ 
the isoscalar electric form
factor is given by the Fourier transform of the baryon density $B_0(r)$
\be
G^0_E (-Q^2) = \frac{1}{2} \int d^3r \; j_0(Qr) B_0(r) \; .
\label{eq:25} \ee
If we assume the existence of a function $\Gamma (t')$ such that the
integral in (\ref{eq:24}) is convergent then spatial density $B_0(r)$ and 
spectral function $\Gamma(t')$ are related through
\be
r B_0(r) = \frac{1}{\pi^2} \int^\infty_{\mu_0} e^{-\mu r} \mu\Gamma
(\mu^2) d\mu
\label{eq:26} \ee
(with notation $t'=\mu^2$), i.e. the function $\mu \Gamma (\mu^2)$ is
the Laplace inverse of $\pi^2 r B_0(r)$. Although
mathematically the problem (\ref{eq:26}) is no less ill-posed than (\ref{eq:24}), models
generally supply analytical information for the behavior of $B_0(r)$
for $r\to 0$ and $r \to \infty$, in which case (\ref{eq:26}) appears better
suited for obtaining the spectral functions.

Specifically, the baryon density $B_0(r)$ for the Skyrme hedgehog is given
by
\be
B_0(r) =-\frac{F' \sin^2\!F}{2 \pi^2 r^2}
\label{eq:27} \ee
and the asymptotic form of the chiral angle $F(r)$ is
\be
 \qquad \;\; F(r) \stackrel{r \to \infty}{\longrightarrow} \frac{A}{r^2} (1 + m_\pi r)
e^{-m_\pi r}
\label{eq:28} \ee
\be
\mbox{and} \qquad F(r) \stackrel{r \to 0}{\longrightarrow} \pi - B r + \cdot \cdot \cdot
\label{eq:29} \ee
where the constants $A$, $B$ are fixed by solving the nonlinear
equation for the soliton profile. The strategy for constructing
$\Gamma(\mu^2)$ then is to approximate the numerically determined
density $rB_0(r)$ by a set of functions with known Laplace inverse such
that the constraints imposed by (\ref{eq:28}), (\ref{eq:29}) 
are fulfilled and (\ref{eq:26}) is
satisfied for all values of $r$.

Commonly, spectral functions are set up as sums of several discrete
'monopoles' at
positions $\mu =\nu_0$ with strengths $a_0$
\be
\Gamma^{(0)} (\mu^2) =\sum  a_0 \delta (\mu^2-\nu^2_0) \; .
\label{eq:30} \ee
Similarly, one may consider 
sums of 'dipoles' at positions $\mu = \nu_1$ with strengths $a_1$
\be
\Gamma^{(1)} (\mu^2) = \sum  a_1 \; \nu_1^2 \delta' (\mu^2 - \nu^2_1) \;.
\label{eq:31} \ee
Their contributions to $rB_0(r)$ according to (\ref{eq:26}) are of the form
$$
2 \pi^2 r B_0(r) =  \Bigl\{  \sum a_0  e^{-\nu_0 r}
+ \sum a_1\left(\frac{\nu_1 r}{2}\right)  e^{-\nu_1 r} \Bigr\} \; .
$$
In addition to (\ref{eq:30}), (\ref{eq:31}) we define 
higher-order pole structures in the spectral function
\begin{eqnarray}
\Gamma^{(2)} (\mu^2)& =& a_2 \left(\nu_2^4 \delta''(\mu^2 - \nu^2_2)
-\frac{1}{2}\nu_2^2\delta'(\mu^2 - \nu^2_2) \right) \;,
\label{eq:32} \\
\Gamma^{(3)} (\mu^2) & = & a_3 \left(\nu_3^6\delta''' (\mu^2 - \nu^2_3)
- \frac{3}{2}\nu_3^4 \delta''(\mu^2 - \nu^2_3)\right) \;,
\label{eq:33} \\
\Gamma^{(4)} (\mu^2) & = & a_4 \left(\nu_4^8\delta'''' (\mu^2 -
\nu^2_4) -3\nu^6_4 \delta'''(\mu^2 - \nu^2_4)
+\frac{3}{4}\nu^4_4 \delta''(\mu^2 - \nu^2_4)
\right) \;, \mbox{~~etc.}
\label{eq:34}
\end{eqnarray}
such that their contributions to $r B_0(r)$ are 
\be
2 \pi^2 r B_0(r) =  \Bigl\{  \sum_{i=0,1,..} a_i
\left(\frac{\nu_i r}{2}\right)^i  e^{-\nu_i r} \Bigr\} \; .
\label{eq:35} \ee
This convenient form proves very efficient for an accurate
representation of numerically obtained functions $r B_0(r)$; we
therefore prefer to use the pole combinations $\Gamma^{(i)}$ instead of
the 'pure multipoles' $\delta^{(i)}$.

Evidently, however, the representation (\ref{eq:35}) does not allow to satisfy the
asymptotic constraint (\ref{eq:28}) which requires that $rB_0(r)$ decreases 
for large $r$ like
\be
2 \pi^2 r B_0(r) \to A^3\frac{N_4(r)}{r^8} e^{-3m_\pi r}
\label{eq:36} \ee
with degree-four polynomial
\be
N_4(r) = (1 + m_\pi r)^2 \Bigl(2 + 2m_\pi r + (m_\pi r)^2\Bigr) \;
. 
\label{eq:37} \ee
In order to correctly reproduce the asymptotic behavior of the
spatial density a term like (\ref{eq:36}) has to be included on the right-hand
side of (\ref{eq:35}). However, to be able to 
satisfy the constraint (\ref{eq:29}) 
for $r \to 0$, the
denominator in (\ref{eq:36}) has to be replaced by a degree-eight polynomial  
\be
D_8(r)=\sum_{i=0}^8 c_i r^i \mbox{~~~~~~~~~~with~~} c_8=1.
\label{eq:38} \ee
Then, the analytical knowledge about $rB_0(r)$ and its derivatives
at $r=0$ fixes the amplitudes of monopoles, dipoles,... in terms of
the coefficients $c_i$.     
In contrast to the discrete structures 
(\ref{eq:30})-(\ref{eq:34}) of the spectral function
which underlie the right-hand side of (\ref{eq:35}) the rational function
$N_4(r)/D_8(r)$ corresponds to a continuous part 
$\Gamma_{\rm cont}(\mu^2)$ of the spectral distribution
$\Gamma(\mu^2)$ which (due to the exponential factor 
$\exp(-3m_\pi r)$) is nonzero only for $\mu \geq 3m_\pi)$. Its form
near threshold is determined by the asymptotics (\ref{eq:28}).

Through an extended analysis of the nonlinear differential equation for
$F(r)$ additional terms in (\ref{eq:28}) for $r \to \infty$ can be
determined which lead to improved numerator polynomials $N_k(r)$ in
(\ref{eq:36}) with $k>4$. Further terms in the 
expansion of $\sin^2F$ in (\ref{eq:27})
carry the exponential factor $\exp(-5m_\pi r)$ and create additional
continuous contributions along the cut for $\mu \geq 5 m_\pi$.
The evaluation of further terms of $F(r)$ in 
(\ref{eq:29}) for $r \to 0$ provide
additional equations to fix more amplitudes in (\ref{eq:35}). In this way
increasingly accurate approximations to $ r B_0 (r)$ could be
constructed. 

There is, however, a severe constraint which has to be imposed on
acceptable denominator polynomials $D_k(r)$:
the Laplace inverse of $N_i/D_k$ (where $k > i$) contains the exponentials
$\exp(\alpha_i \mu)$ where $\alpha_i$ are the roots of $D_k$. Therefore
the integral in (\ref{eq:26}) converges only for 
$r >$ Re$\:\alpha_{\max}$, where
$\alpha_{\max}$ is the root with the largest positive real part.
Consequently all roots of acceptable denominator polynomials must lie
in the left complex half plane. (For simple roots the imaginary axis is
still acceptable).

The fact that the "data", the numerically obtained function $B_0(r)$,
is known with great precision allows to create very satisfactory fits 
by including a sufficient number of
terms at different positions $\nu_i$ in the sum on the r.h.s. of
\be
2 \pi^2 r B_0(r) = A^3\frac{N_4(r)}{D_8(r)}e^{-3m_\pi r}+
\Bigl\{ \sum_{i=0,1,..}  a_i\left(\frac{\nu_i r}{2}\right)^i e^{-\nu_i r}
\Bigr\} \; 
\label{eq:39} \ee
and constrain their amplitudes $a_i$ through boundary conditions at $r = 0$
\be
r B_0(r)|_0 = 0 \; ,\mbox{~~~~} (r B_0(r))'|_0 = B_0(0) \; , 
\mbox{~~~~} (rB_0(r))''|_0 =0 \; ,
\mbox{~~~~}(r B_0(r))'''|_0 =3 B_0''(0)\; ,
\label{eq:40} \ee
or, explicitly,
\begin{eqnarray}
0&=&2 A^3\: \frac{1}{c_0} + a_0 
\label{eq:41} \\
2 \pi^2 B_0 (0)&=&- 2 A^3\: \frac{c_1}{c^2_0} - a_0 \nu_0 + 
\frac{1}{2}a_1\nu_1 \nonumber\\
0&=&- 4 A^3\: \frac{m^2_\pi c^2_0 + c_0 c_2 - c_1^2}{c^3_0} + a_0 \nu^2_0 -
a_1\nu_1^2 + \frac{1}{2}a_2 \nu^2_2  \nonumber\\
6\pi^2 B_0''(0) & = & 2A^3\frac{3m_\pi^3c_0^3+6m_\pi^2c_0^2c_1
+12c_0c_1c_2-6c_0^2c_3-6c_1^3}
{c_0^4}-a_0\nu_0^3+\frac{3}{2}a_1 \nu_1^3-\frac{3}{2}a_2\nu_2^3
+\frac{3}{4}a_3\nu_3^3. \nonumber    
\end{eqnarray}
The first of these equations shows that
at least one discrete monopole is necessary to allow for 
$rB_0(r)|_0=0$, and that its discrete monopole strength $a_0$,
however, is completely
compensated by the continuous part of the spectral function, because we
have from (\ref{eq:39}) and (\ref{eq:26}) that
\be
\int \Gamma_{\rm cont}(\mu^2) d\mu^2 =  A^3\frac{N_4}{D_8}|_{r=0}
=\frac{2A^3}{c_0}.
\label{mono}
\ee
So, evidently, the origin of the dipole nature of the form factor lies
in the fact that the density $B_0(r)$ has a finite limit at $r$=0.

Due to the fact that the set of functions which appear on the r.h.s. of
(\ref{eq:39}) is neither orthogonal nor complete, there remain severe
ambiguities in their selection. From the physical point of view we
might prefer to include only a set of monopoles at different positions.
This, however, proves very inefficient for an accurate fit. Including
more terms in (\ref{eq:39}) leads to improved fits; however, as usual, it does
not make sense to include many more terms, because different fits of
comparable (improved) accuracy may differ appreciably in their
resulting distribution of multipole strengths. We find the best
results (with respect to stability and accuracy) by including the four
terms with $i=0,..,3$ in (\ref{eq:39}), i.e. just one term for each
value of $i$ at positions $\nu_i$ with the four amplitudes $a_i$
fixed through the four relations (\ref{eq:41}). 
The resulting pole structure can only be interpreted as
an effective discrete representation of some undetermined underlying
strength distribution. As is evident from (\ref{eq:41}), location and strength
of these effective pole structures are in close connection to the
additional continuous part which is severely constrained through the
asymptotic behavior of the spatial density.
Thus (\ref{eq:39}) with $i=0,..,3$ appears as a simple and natural form
which satisfies the asymptotics for $r\to 0$ and $r\to \infty$ and 
allows for a very accurate fit of $rB_0(r)$.

Allowing the roots of $D_8$ to vary freely in the left
complex half plane leads to two distinct double roots and their complex
conjugates, i.e.
\be
D_8 = |(r-\alpha_1)^2 (r-\alpha_2)^2|^2.
\label{eq:42} \ee
Imposing this form for $D_8$ leaves two complex roots $\alpha_1$,
$\alpha_2$ and four pole positions $\nu_i$, $(i=0,..,3)$ as
parameters for the fit (\ref{eq:39}). We find that the 
position of the roots $\alpha_1$, $\alpha_2$ is quite stable with
respect to different pole combinations, and the smallest $\chi^2$ is
obtained for the combination given in (\ref{eq:39}).

With (\ref{eq:42}) the decomposition of $N_4/D_8$ into partial fractions is
\be
\frac{N_4(r)}{D_8(r)} = \frac{\gamma_1}{r - \alpha_1} +
\frac{\gamma_{11}}{(r - \alpha_1)^2} + \frac{\gamma_2}{r - \alpha_2} +
\frac{\gamma_{22}}{(r-\alpha_2)^2} + c.c.
\label{eq:43} \ee
where the coefficients $\gamma_i$ satisfy relations like
\begin{eqnarray}
\mbox{Re} \sum_{i=1,2} \gamma_i & = & 0 \;  \nonumber \\
\mbox{Re} \sum_{i=1,2} (\gamma_{ii} +\alpha_i\gamma_i) & = & 0 \; 
 \nonumber \\
\mbox{Re} \sum_{i=1,2} (2\alpha_i\gamma_{ii} +\alpha_i^2\gamma_i) 
& = & 0 \; 
\label{eq:44}
\end{eqnarray} 
and so forth, which imply that the Laplace inverse $\gamma(\mu)$ of
$N_4/D_8$
\be
\gamma(\mu)=2 \mbox{Re}\sum_{i=1,2} (\gamma_i +\gamma_{ii}\: \mu)
e^{\alpha_i \mu}
\label{eq:45} \ee
satisfies at threshold $\mu=0$
\be
\gamma(0)=0;~~~~\gamma'(0)=0;~~~~\gamma''(0)=0;~~~~
\gamma'''(0)= m_\pi^4;~~~~\mbox{etc.}
\label{eq:46} \ee 
(These relations, of course, follow also quite generally by 
comparing the coefficients for positive powers of $r$ for $r \to 
\infty$ in the definition of $\gamma(\mu)$
\be 
N_i(r)=D_k(r)\int_0^\infty e^{-\mu r}\gamma(\mu)d\mu
\label{eq:47} \ee
for polynomials $N_i(r)$ and $D_k(r)$.)
From (\ref{eq:26}) and (\ref{eq:39}) the continuous part of $\Gamma(\mu^2)$ then is given
by
\be
\Gamma_{\rm cont}(\mu^2)=\frac{A^3}{2 \mu}\gamma(\mu-3m_\pi)
\label{eq:48} \ee
for $\mu\geq 3 m_\pi$. At threshold $\mu^2 \to (3 m_\pi)^2$ its
behavior is fixed by (\ref{eq:44}) as 
\be
\Gamma_{\rm cont}(\mu^2) \to \frac{A^3}{6^5}(\mu^2-(3m_\pi)^2)^3.
\label{eq:49} \ee
It should be noted that this onset of the spectral density at threshold
is completely determined by the form of the baryon density (\ref{eq:27}) and
the asymptotics of the chiral angle (\ref{eq:28}) and is
independent of the discrete structures added in (\ref{eq:39}). It cannot be 
obtained in parametrizations of spectral functions in terms of discrete
pole structures only, but the existence of this continuous part affects 
positions and amplitudes of poles through equations like (\ref{eq:41}).

Altogether, the isoscalar electric spectral function is obtained in the
form
\be
\Gamma^{I=0}(t)=\Gamma_{\rm cont}(t)
+\sum_{i=0}^3\Gamma^{(i)}(t)
\label{eq:50} \ee
where the continuous part is given by 
(\ref{eq:48}) with (\ref{eq:45}) and the amplitudes
of the discrete pole structures (\ref{eq:30})-(\ref{eq:33}) 
are fixed through (\ref{eq:41}).

For $\mu^2 \gg (3m_\pi)^2$ the continuous part
$\Gamma_{\rm cont}(\mu^2)$ is dominated by the exponential
$\exp \alpha_i \mu$ corresponding to the root 
$\alpha_i = -\varepsilon_i \pm i \varphi_i$  of $D_8$
with the smallest (absolute) value of its
real part $\varepsilon_i$, i.e. $\Gamma_{\rm cont}$ oscillates 
with slowly decreasing amplitude. 
It turns out that the imaginary part $\varphi_i$ (i.e. the
period of the oscillations) is sharply fixed by the fit, while
variations of the real part 
$-\varepsilon_i$ within a limited range affect the
quality of the fit only very little. (By putting a constraint on
$\varepsilon_i$ the oscillations thus can be damped without
severe consequences for the fit).

As an example we present the result of such an analysis for the 
standard Skyrme model (with $f_\pi$ = 93 MeV and $e$ = 4.25).
A typical fit leads to the (two-fold) roots 
(in units of (inverse) $m_\varrho$ = 770 MeV)
\be
\alpha_1 = - 0.725 \pm i\:1.505 \quad ; \quad \alpha_2 = - 0.5 \pm i
\: 1.823 \mbox{~~~~~}[m_\varrho^{-1}] \; .
\label{eq:51} \ee
Here the real part of $\alpha_2$ has been kept fixed at -0.5. If 
$\varepsilon_2$ is allowed to move freely it will slowly approach 
$\varepsilon_2$=0 with only marginal improvement of the fit and little
change in pole positions. The choice (\ref{eq:51}) efficiently cuts down
the oscillations in the continuous part of the spectral function 
for large $\mu$ (see fig.3).
The pole locations and strengths for the same fit are listed in table I.

\begin{table}[h]
\begin{tabular}{|c|cccc|}
$\Gamma^{(i)}$ & $i = 0$ & $i = 1$ & $i = 2$ & $i=3$~~~~~~~~~~~  \\
\hline
\rule{0mm}{5mm}
$\nu_i\;\;[m_\varrho]\;$~~~ 
&  1.61  &  5.14  &  1.93  &  2.09~~~~~~~~~ \\
~$a_i \;\;[m_\varrho^2]\;$~~~ & -8.80 & 0.47 & 15.99 & -4.89~~~~~~~~~~
\end{tabular}
\caption[]{Positions and strengths of the discrete structures for the
isoscalar electric spectral function (in the soliton
restframe)}
\end{table}

A very stable feature of this result is the discrete monopole
strength $\Gamma^{(0)} (t) $ near 1.6 $m_\varrho$. 
Of course, this discrete monopole strength is completely compensated 
by the continuous part $\Gamma_{\rm cont}(t)$
which is very small near $t\approx m_\varrho^2$ but has a first
pronounced maximum near $(2.5 m_\varrho)^2$ (see fig.3).
Also this feature of $\Gamma_{\rm cont}(t)$ is very stable. 
There is no discrete low-lying dipole strength.
But $\Gamma^{(2)}$ contains dipole strength and 
strong quadrupole strength near 1.5 GeV. 
Together with some octupole strength this may indicate broader underlying 
structures, although the details are certainly peculiar 
to our specific choice of allowed pole structures in the ansatz (\ref{eq:39}).  
Altogether it appears that in this way we obtain a quite reliable
picture of the spectral function which underlies the isoscalar
electric form factor of the Skyrme model up to about 2 - 3 $m_\varrho$.

\subsection{Isovector electric $G_E^1(t)$}

For the isovector electric form factor 
\be
G^1_E(-Q^2) = \frac{1}{2} \int d^3 r \; j_0 (Qr) B_1(r)
\label{eq:52} \ee
the baryon density $B_0(r)$ in
(\ref{eq:25}) and (\ref{eq:26}) is replaced by the (isorotation) inertia density
\be
B_1(r) = \frac{2}{3 \Theta} \sin^2 F \Bigl(f^2_\pi + \frac{1}{e^2}(F'^2
+ \frac{\sin^2 F}{r^2}) \Bigr) \; ,
\label{eq:53} \ee
normalized by the moment of inertia $\Theta$ such that $\int B_1(r) d^3
r = 1$. The asymptotic constraint (\ref{eq:28}) then suggests a rational function
$N_2/D_4$ on the right-hand side in the representation of $B_1(r)$
(corresponding to (\ref{eq:39}))
\be
2 \pi^2 B_1(r) = f_1 \frac{N_2 (r)}{D_4 (r)} e^{-2m_\pi r} +
\frac{1}{r} \Bigl\{ \sum a_0 e^{-\nu_0 r} + \sum a_1\frac{\nu_1 r}{2} 
e^{-\nu_1 r} + \cdot \cdot \cdot \Bigr\}
\label{eq:54} \ee
with
\begin{eqnarray}
f_1 & = &  \frac{4 \pi^2}{3 \Theta} f^2_\pi A^2 \nonumber \\
\rule{0mm}{5mm}
N_2(r) & = & ( 1 + m_\pi r)^2 \nonumber \\
\rule{0mm}{5mm}
D_4 & = & \sum^4_{i = 0} c_i r^i \quad \mbox{~~~~with~~} \; c_4 = 1 \; .
\label{eq:55}
\end{eqnarray}
The boundary conditions at $r = 0$ are 
\be
B_1(0) = 0 ;\;\; B'_1(0) = 0 ; \;\; 
B''_1(0) = \frac{4}{3 \Theta} B^2 (f_\pi^2 + \frac{2}{e^2}B^2) ; 
\;\; B'''_1(0) = 0 ;
\label{eq:56} \ee
These conditions exclude discrete monopole strength. It turns out
that at least four discrete structures are needed for satisfactory fits.
Again we find that by far the best result is obtained if we allow for
just one term of each power of $r$ in (\ref{eq:54}), i.e. we use
\be
2 \pi^2 B_1(r) = f_1 \frac{N_2}{D_4} e^{-2m_\pi r} +
\frac{1}{r} \Bigl\{ \sum^4_{i=1}  a_i(\frac{\nu_i r}{2})^i e^{-\nu_i r}
\Bigr\} \; ,
\label{eq:57} \ee
where the four amplitudes $a_i$ are fixed through relations (\ref{eq:56}), which
read explicitly
\begin{eqnarray*}
0 & = & f_1\frac{1}{c_0} + a_1 \frac{\nu_1}{2} \\ 
\rule{0mm}{5mm}
0 &  = & -f_1 \frac{c_1}{c_0^2} - a_1 \frac{\nu^2_1}{2} + a_2
\frac{\nu^2_2}{4} \\
\rule{0mm}{5mm}
2 \pi^2 B''_1(0)  & = &  - 2 f_1 \frac{(m^2_\pi c_0 + c_0 c_2 -
c_1^2)}{c^3_0} + a_1 \frac{\nu^3_1}{2} - a_2 \frac{\nu^3_2}{2} + 
 a_3 \frac{\nu^3_3}{4} \\
\rule{0mm}{5mm}
0 & = & f_1 \frac{4 m^3_\pi c^3_0 + 6 m^2_\pi c_1 c^2_0 - 6 c^2_0 c_3 +
12 c_0 c_1 c_2 - 6 c^3_1}{c^4_0} 
 -   a_1 \frac{\nu^4_1}{2} + a_2 \frac{3 \nu^4_2}{4} - a_3 \frac{3
\nu^4_3}{4}  + a_4 \frac{3 \nu^4_4}{8} 
\end{eqnarray*}

The two complex roots $\alpha_1, \alpha_2$ of the denominator polynomical
$D_4 (r)$ of degree 4
\be
D_4(r) = |(r-\alpha_1)(r-\alpha_2)|^2
\label{eq:58} \ee
together with the pole positions $\nu_i$ ($i$=1,..,4) provide 8
real parameters for a very accurate fit. Again the roots $\alpha_i = -
\varepsilon_i \pm i \varrho_i$ are restricted to the left complex half
plane.

It is instructive to notice that (in contrast to the isoscalar case)
the different asymptotic behavior of $B_1 (r)$ for $r \to \infty$
causes a discontinuity of the spectral function at threshold $\mu^2 \to
(2m_\pi)^2$:
the decomposition of $N_2/D_4$ into partial fractions
\be
\frac{N_2(r)}{D_4(r)} = \frac{\gamma_1}{r - \alpha_1} +
\frac{\gamma_2}{r - \alpha_2} + c.c.
\label{eq:59} \ee
with relations like
\begin{eqnarray}
\mbox{Re} \sum_{i =1,2} \gamma_i & = & 0,  \nonumber \\
2 \mbox{Re} \sum_{i =1,2} \alpha_i \gamma_i & = & m^2_\pi,  \nonumber \\
 \mbox{Re} \sum_{i =1,2} \alpha_i (\alpha_i \gamma_i - m^2_\pi) & = & m_\pi 
\label{eq:60}
\end{eqnarray}
shows that the Laplace inverse $\gamma(\mu)$ of $r N_2/D_4$
\be
\gamma(\mu) = 2 \mbox{Re} \sum_{i =1,2} \alpha_i \gamma_i e^{\alpha_i \mu}
\quad (\mu \geq 0)
\label{eq:61} \ee
as $\mu \to 0$ takes on the value 
\be
\gamma(0) = m^2_\pi  
\label{eq:62} \ee
with a slope of
\be
\gamma'(0) = 2 m_\pi ( 1 + m_\pi \mbox{Re} \sum_{i=1,2} \alpha_i) \; .
\label{eq:63} \ee
This slope is positive and, due to the small value of $m_\pi$, 
it is not very sensitive to small variations in 
Re$(\alpha_1+\alpha_2)$. 

Altogether, we obtain for the isovector electric spectral function 
\be
\Gamma^{I=1}(t')=\Gamma_{\rm cont}(t')
+\sum_{i=1}^4 \Gamma^{(i)}(t')
\label{eq:64} \ee
where the continuous part is given by 
\be
\Gamma_{\rm cont}(\mu^2) = \frac{f_1}{2 \mu} \gamma(\mu - 2
m_\pi) \; 
\label{eq:65} \ee
with (\ref{eq:61}) and the amplitudes $a_i$
of the discrete pole structures (\ref{eq:31})-(\ref{eq:34}) 
are fixed through (\ref{eq:56}).

As before it turns out that the quality of the fit is not very 
sensitive to variations of $\varepsilon_2$ within 
$(0.2 < \varepsilon_2 < 0.4)$. Keeping $\varepsilon_2$ fixed at 0.3
the positions of the roots of $ D_4$ are determined as 
\be
\alpha_1 = -0.550 \pm i\: 0.215 \quad ; \quad \alpha_2 = -0.3 \pm i\:0.626
\quad ; \quad [m^{-1}_\varrho]
\label{eq:66} \ee
with pole locations and amplitudes given in table II.

\begin{table}[h]
\begin{tabular}{|c|cccc|}
$\Gamma^{(i)}$ & $i = 1$ & $i = 2$ & $i = 3$ & $i = 4$~~~~~~~~~ \\ \hline
\rule{0mm}{5mm}
$\nu_i \;\;[m_\varrho]\;$~~~ 
&  4.39  &  0.72  & 2.60  &  2.92~~~~~~~~~~~ \\ 
~$a_i \;\;[m_\varrho^2]$~~~ 
& -7.59 & 0.009 & -5.59 & 3.49~~~~~~~~~~~ 
\end{tabular}
\caption[]{Positions and strengths of the discrete structures for the
isovector electric spectral function (in the soliton
restframe)}
\end{table}

The stable feature of the resulting isovector spectral function is that there
is no significant low-lying discrete strength, while the continuous
part rises steeply from its finite limit 
\be
\Gamma_{\rm cont}(\mu^2 \to (2m_\pi)^2) \to  \frac{f_1 m_\pi}{4}
\label{eq:67} \ee
at threshold to a pronounced 
maximum near $1.5 m_\varrho$ (see fig.4). Again the discrete structures above
2$ m_\varrho$ mainly reflect our specific choice (\ref{eq:57}).

\subsection{Magnetic form factors $G^0_M(t)$ and $G^1_M(t)$}

In the purely pionic Skyrme model, the normalized 
magnetic isoscalar and isovector
form factors are given by $(t = -Q^2)$
\be
G^0_M(t)  =  \frac{3}{2 r_B^2} \int d^3 r \; \frac{j_1 (Q r)}{Qr} r^2 B_0
(r) \mbox{~~~~~~~~~~}
G^1_M(t)  =  \frac{3}{2} \int d^3r \;\frac{j_1 (Q r)}{Qr} B_1 (r) \; .
\label{eq:68} \ee
Thus, with the convenient supplementary definition
\be
\tilde G^1_E(t)  \equiv   \frac{1}{2} \int d^3 r \; j_0 (Q r) \frac{
B_1(r)}{r^2} 
\label{eq:69} \ee
we have
\be
G^0_M(t)  =  \frac{6}{r_B^2}\: \frac{\partial}{\partial t} G^0_E (t) , 
\mbox{~~~~~~~~~~}
G^1_M(t)  =  6 \frac{\partial}{\partial t} \tilde G^1_E (t)
 \; .
\label{eq:70} \ee
and, correspondingly, for the spectral functions $\Gamma(t')$
\be
\Gamma^0_M(t')  =  \frac{6}{r_B^2} \;\frac{\partial}{\partial t'}
\Gamma^0_E (t') \; ,
\mbox{~~~~~~~~~~}
\Gamma^1_M(t') =  6 \frac{\partial}{\partial t'}
\tilde \Gamma^1_E (t')
\label{eq:71} \ee
for timelike $t' = \mu^2$.

The Laplace transformation (\ref{eq:26}) relates $\Gamma^1_E(\mu^2)$ and
$\tilde\Gamma^1_E (\mu^2)$ by
\be
\mu \tilde \Gamma^1_E (\mu^2) = \int^\mu_{\mu_0} \int^{\nu'}_{\mu_0}
\nu \Gamma^1_E (\nu^2) \; d \nu \; d \nu'
\label{eq:72} \ee
(with $\mu_0 = 2 m_\pi$) .

For the continuous part of the isoscalar magnetic spectral function we
obtain from (\ref{eq:48}) and (\ref{eq:71})
\be
\Gamma^0_{M\; cont} (\mu^2) \to \frac{3}{r_B^2} \frac{1}{\mu}
 \frac{\partial}{\partial \mu} \frac{A^3}{2\mu}\gamma(\mu-3m_\pi)
\label{eq:73} \ee
with $\gamma(\mu)$ given by (\ref{eq:45}). At threshold $\mu \to 3m_\pi$
$\Gamma_M^0$ rises like
\be
\Gamma^0_{M\; cont} (\mu^2) \to \frac{3}{r_B^2} \frac{A^3}{6^4}
\Bigl(\mu^2 - (3m_\pi)^2\Bigr)^2 \; .
\label{eq:74} \ee

The discrete monopole structure near $\mu \sim 1.6 m_\varrho$ which appeared
as a stable feature in $\Gamma^0_E(\mu^2)$ is turned into a dipole by
the derivative in (\ref{eq:71}). Therefore $\Gamma^0_M$ contains no discrete
monopole strength.

From (\ref{eq:71}), (\ref{eq:72}), (\ref{eq:61}) 
and (\ref{eq:65}) we obtain the continuous part of the
isovector magnetic spectral function
\be
\Gamma^1_{M \; cont} (\mu^2) 
= 3 \frac{f_1}{\mu^3} \; \mbox{Re}
\sum_{i=1,2} \frac{\gamma_i}{\alpha_i} \left( 1 + (\alpha_i \mu - 1)
e^{\alpha_i (\mu - 2 m_\pi)}\right) \; .
\label{eq:75} \ee
These continuous parts of the magnetic spectral functions are 
displayed by the dashed lines in figs.3,4.

\section{Coupling to vector mesons}
We have accounted for the influence of vector mesons on the form
factors by multiplying the purely pionic parts $G^{\pi}(t)$ 
with 
\be
\Lambda(t)=\lambda D_V(t) +1-\lambda
\label{eq:76} \ee
where $D_V(t)$ is just the vector meson propagator. To illustrate its
influence on the spectral functions we consider the isovector electric
case where the pionic part of the spectral function is given by (\ref{eq:64})
\be
\Gamma^{\pi}(t')=\Gamma_{\rm cont}(t')
+\sum_{i=1}^4 \Gamma^{(i)}(t')
\label{eq:77} \ee
with the continuous part $\Gamma_{\rm cont}$ shown in fig.4
and positions and amplitudes of the discrete part $\sum \Gamma^{(i)}$
listed in table II.

If we denote the imaginary part of the $\varrho$-propagator
${\cal I}m \:D_V = \Gamma^{\varrho}(t')$ we
have for the combined form factor (for $\lambda$=1)
\be 
G(t) = D_V(t) G^{\pi}(t)= 
\frac{1}{\pi^2}\int \;\frac{\Gamma^{\varrho}(t')}{t'-t} \;
\frac{\Gamma^{ \pi}(t'')}{t''-t}\:dt'\:dt'' .
\label{eq:78} \ee
For the spectral function $\Gamma(t')$ of $G(t)$ this implies
\be
\Gamma(t')=\frac{1}{\pi}\left(\Gamma^{\varrho}(t')\;{\cal P}\!\!\!\int
\frac{\Gamma^{ \pi}(t'')}{t''-t'}dt'' 
\;+\; \Gamma^{ \pi}(t')\;{\cal P}\!\!\!\int
\frac{\Gamma^{\varrho}(t'')}{t''-t'}dt''\right)
\label{eq:79} \ee
where ${\cal P}$ denotes the principal value. The first integral 
gives the resulting strength distribution of the vector-meson pole.
It receives contributions from the continuous and the discrete part of
$\Gamma^{ \pi}$ which are slowly varying functions of $t'$ within the
resonance region. At $t'$=$m_V^2$ we obtain the values
\be
{\cal P}\!\!\!\int \frac{\Gamma_{\rm cont}(t'')}{t''-m_V^2}dt'' = 0.947,
~~~~~~~~~~~\int\frac{\sum\Gamma^{(i)}(t'')}{t''-m_V^2}dt'' = 1.192
\label{eq:80} \ee
so that the total strength of the vector-meson pole is increased by a factor 
of 2.14.

The second integral in (\ref{eq:79}) modifies the pionic spectral function. 
The discrete multipole structures $\Gamma^{(i)}$ in
$\Gamma^{ \pi}$ of course remain discrete, but receive
additional lower multipoles at the same position. (Only monopoles
remain monopoles, with modified strength). 
Of particular interest is the modification in the 
continuous part of the spectral function: The $t'$-dependence
of the second integral in (\ref{eq:79}) cuts down the tails
in $\Gamma_{\rm cont}(t')$ for larger values of $t'$
and creates a sign change near the vector-meson pole. 
The result is shown in fig.5: For the demonstrative purpose
we use for the $\varrho$-pole the imaginary
part (dotted line in fig.5)
\be 
\Gamma^{\varrho}(t')=\frac{m_V^3\varepsilon}{(m_V^2-t')^2+m_V^2\varepsilon^2}
\label{eq:81} \ee
with the experimental full width $\varepsilon$=150 MeV (and a smooth cutoff
near the two-pion threshold). The dashed line in fig.5 
is the continuous pionic part $\Gamma_{\rm cont}$ from 
(\ref{eq:65}) and fig.4. 
The dash-dotted line shows the continuous 
spectral function $\Gamma_E^{(T=1)}(t')$ 
as resulting from (\ref{eq:79}). The increased strength of the resonance due
to the integrals (\ref{eq:80}) is evident. Furthermore the figure
clearly shows that the interplay between the vector meson and the solitonic 
pion cloud leads to appreciable enhancement of the resonance
structure on the low-energy side and a corresponding depletion on the 
high-energy side. This is strongly reminiscent of the Frazer-Fulco
unitarity constraint which enforces a similar behavior
for the isovector spectral functions above the two-pion 
threshold~\cite{Hoe76,meisdrech}.
Notable from the figure is also the finite limit (\ref{eq:67}) of 
$\Gamma_E^{(T=1)}(t'$=$4m_\pi^2) = f_1 m_\pi /4$
at threshold. 

The actual value of $\lambda$ used in sect.1 for the form factors was
$\lambda$=0.75. Therefore the total spectral functions are obtained as
the corresponding superpositions of $\lambda\Gamma(t')$ from (\ref{eq:79}) and
the purely pionic $(1-\lambda)\Gamma^\pi(t')$ of (\ref{eq:64}). 
This final result
for $\Gamma_E^{(T=1)}(t')$ is shown by the full line in fig.5.

\section{Relativistic corrections}
The Lorentz boosts (\ref{eq:16}) and (\ref{eq:17}) which relate 
the form factors as evaluated in 
the soliton restframe and in the Breit frame can be directly
transferred to their respective spectral functions.
If we denote by $\Gamma^{(nr)}(t')$ the spectral function as
evaluated in the soliton restframe, we have from (\ref{eq:16})
\begin{eqnarray}
G_E(t) & = & \frac{1}{\pi} \int^\infty_{t_0}
\frac{\Gamma^{(nr)}_E(t')}{t'-t/\gamma^2}  \; dt'  \nonumber \\
& = & \frac{1}{\pi} \int^{4M^2}_{\tilde t_0}
\frac{\Gamma^{(nr)}_E(\tilde t/\tilde \gamma^2)}{\tilde t - t} \; d\tilde t
\;\; + \;\; G^{(nr)}_E (-4M^2)
\label{eq:82}
\end{eqnarray}
with
\be
\tilde \gamma^2 = 1 - \frac{\tilde t}{4 M^2} \quad \mbox{for timelike}
\quad \tilde t > 0 \; .
\label{eq:83} \ee
Thus, with
\be
G_E (t) = \frac{1}{\pi} \int^{4M^2}_{\tilde t_0} \frac{\Gamma_E (\tilde
t)}{\tilde t - t} \;d \tilde t \;\; + \quad \mbox{const.}
\label{eq:84} \ee
we have
\be
\Gamma_E(\tilde t) = \Gamma^{(nr)}_E (\tilde t / \tilde \gamma^2) \; .
\label{eq:85} \ee
The additional constant which appears in (\ref{eq:82}) and
(\ref{eq:84}) is the undesirable consequence
of (\ref{eq:16}) for $t \to \infty$
\be
G_E (t \to \pm \infty) \to  G^{(nr)}_E (-4M^2)
\label{eq:86} \ee
which we have met already in (\ref{eq:18}).

The integration interval in (\ref{eq:84}) is determined by the threshold of
$\Gamma^{(nr)}_E$, i.e. $\tilde t /\tilde\gamma^2 = t_0$, or
\be 
\tilde t_0= \frac{t_0}{1+t_0/(4M^2)}.
\label{eq:87} \ee
The upper limit is the physical threshold for soliton-antisoliton
creation which follows from (\ref{eq:85}) for $\tilde \gamma^2 \to 0$.

For the magnetic form factors the boost to the Breit frame (\ref{eq:17})
leads to
\be
G_M(t) = \frac{1}{\pi} \int^{4M^2}_{\tilde t_0} \frac{\Gamma_M(\tilde
t)}{\tilde t - t} d \tilde t
\label{eq:88} \ee
with
\be
\Gamma_M(\tilde t) = \frac{1}{\tilde \gamma^2} \Gamma_M^{(nr)} (\tilde
t/\tilde \gamma^2) \; .
\label{eq:89} \ee
In this case there remains no additional constant term in (\ref{eq:88}) 
such that $G_M(t)$ satisfies $G_M (t \to \pm \infty) \to 0$.  The factor
$1/\tilde \gamma^2$ in the relativistically corrected magnetic spectral
functions diverges as $\tilde t \to 4 M^2$. This amplifies the
oscillations of the non relativistic $\Gamma_M^{(nr)} (t')$ for $t' \gg
m^2_\varrho$ which are not well determined by the fit (through
$\varepsilon_2$) and thus should not be taken too seriously.

The requirement that the threshold $\tilde t_0$ in (\ref{eq:84}) should 
appear at multiples of the physical pion mass $\tilde t_0=(n \:m_\pi)^2$  
($n$ = 2 or 3 for isovector or isoscalar spectral functions)
can only be met if in the soliton restframe an effective
pion mass $\mu_\pi$ is employed:
\be
t_0=(n\: \mu_\pi)^2= \frac{(n\: m_\pi)^2\;\; 4M^2}{4M^2-(n\: m_\pi)^2}.
\label{eq:90} \ee
This is an extension to timelike $t_0$ and $\tilde t_0$ of the
usual 'reduced mass' formula 
\be
(n \: \mu_\pi)^2= \frac{(n\: m_\pi)^2 \;\; 4M^2}{4M^2+(n\: m_\pi)^2}
\label{eq:91} \ee
for spacelike momentum transfer. This effect, however, is very small.

The main effect of the transformations (\ref{eq:85}) and (\ref{eq:89}) 
is that the point
$t'=\infty$ is mapped onto the threshold $\tilde t = 4 M^2$, i.e.
the nonrelativistic spectral functions  are squeezed into the 
finite interval $\tilde t_0 < \tilde t < 4 M^2$. For the continuous
parts of the pionic spectral functions this is shown in the lower
sections of figs.3,4.
Of course, the low-energy region is almost unaffected, especially
the region around the vector meson resonance (after
inclusion of the $\varrho$-pole) remains essentially as shown in fig.5. 
But the positions $\nu_i$ of the higher discrete structures listed in tables
I,II are shifted to $\tilde \nu_i=\nu_i/\sqrt{1+\nu_i^2/4M^2}$. 

\begin{table}[h]
\begin{tabular}{|c|cccc|cccc|}
&&isoscalar&&&&isovector&&\\
\hline
  & $i = 0$ & $i = 1$ & $i = 2$ & $i = 3$
& $i = 1$ & $i = 2$ & $i = 3$ & $i = 4$ \\ \hline
\rule{0mm}{5mm}
$~\nu_i \;\;[m_\varrho]\;$~~ 
& $\;$ 1.61  $\;$  & $\;$5.14  $\;$ & 
$\;$ 1.93  $\;$  & $\;$ 2.09  $\;$ 
& $\;$ 4.39  $\;$  & $\;$ (0.72)  $\;$ & 
$\;$ 2.60  $\;$  & $\;$ 2.92  $\;$ \\ 
~~$\tilde \nu_i \;\;[MeV]$~~ 
& $\;$ 1034 $\;$  & $\;$ 1695 $\;$ & 
$\;$ 1165 $\;$  & $\;$ 1221 $\;$ 
& $\;$ 1640 $\;$  & $\;$  $\;$ & 
$\;$ 1369 $\;$  & $\;$ 1440 $\;$ \\ 
\end{tabular}
 
\caption[]{Positions of the calculated discrete structures: $\nu_i$, 
in the soliton restframe (taken from tables I,II),  and $\tilde \nu_i$,
in the Breit frame (as obtained from 
$\tilde \nu_i=\nu_i/\sqrt{1+\nu_i^2/4M^2}$ with $M$=$M_N$).}
\end{table}

If we here use for $M$ the physical nucleon mass $M_N$=940 MeV in order
to relate these shifted positions to the actual physical 
$N$-$\bar N-$threshold we obtain the values listed in table III. Although
we cannot really expect any close correspondence to the actually observed
isoscalar ($\Phi$(1020), $\omega$(1420), $\omega$(1600), $\Phi$(1680)) and 
isovector ($\varrho$(1450), $\varrho$(1700)) resonances it is satisfactory
to find the calculated structures in this energy range. (The isovector
structure at 0.72 $m_\varrho$ appears with almost vanishing strength 0.009 
in table II and may indicate that the continuous part expects small
modifications near 5 $m_\pi$.)

\section{Summary}
In view of the widespread prejudice that baryons considered as solitons
in effective mesonic theories provide 
only poor approximations for the e.m. nucleon form factors within a
very limited range of $Q^2$ 
we demonstrate here that even in the most simple soliton model 
the essential features of form factors observed over three orders 
of magnitude in momentum transfer are naturally reproduced.  
The analysis shows that in order to achieve such a result it is
necessary to incorporate three basic ingredients: 
an extended object, partial coupling to vector mesons, and 
relativistic recoil corrections. We have used here the standard
skyrmion as the extended object, and (for simplicity)
the same vector meson propagator
in both isospin channels. Apart from the Skyrme constant $e$ which is
adjusted to the Delta-nucleon mass split the model contains only one
parameter of direct relevance to the form factors: the amount
$\lambda$ which measures the admixing of photon-vector meson coupling.
We use it to bring the e.m. proton radii close to their experimental
values. This is sufficient to keep the form factors $G_{E,M}^p$ and
$G_M^n$ very close to the standard dipole up to almost 1 GeV/c.

Naturally, the typical shape above $Q^2\sim M_N^2$ is sensitive to the
recoil corrections. We employ the boost transformation to the Breit
frame suggested by Ji~\cite{Ji91} which pulls the point $Q^2=4M^2$ (in
the restframe) to $Q^2 =\infty$ (in the Breit frame). Thus the
behavior of the form factors for $Q^2\to\infty$ is, of course, very
sensitive to the value of $M$, the kinematical mass of the
particle represented by the soliton. In tree approximation $M$ is the
classical soliton mass, but it is known to be subject to large 
quantum corrections. The boost to the Breit frame has the 
undesired feature that it violates superconvergence unless $4M^2$
coincides with a zero in $G_M$ (or a double zero in $G_E$). It should 
be noted, however, that the boost transformation is not an exact
result, but its derivation relies on neglecting all commutators between
center-of-mass and total momentum~\cite{Ji91}. Thus it appears as a reasonable 
procedure to subtract the very small undesired terms from the boosted
form factors. We have shown here that up to 10 (GeV/c)$^2$ their effect
is qualitatively unimportant and the typical features observed
in $G_M^p$ (which is the only form factor known sufficiently well up to
high $Q^2$) clearly emerge from the model. Due to these difficulties 
connected with the treatment of recoil corrections the predictive 
power at tree level is poor for the high-$Q^2$ behavior. This 
flexibility is unfortunate because it prevents definite conclusions
and conceals possible evidence for importance of QCD effects.
From this point of view it would be highly  
desirable to learn more about loop and rotational corrections to the
form factors.

The continuation to timelike $t$ leads to quite stable results for the
spectral functions in the regime from the 2- or 3-pion threshold to
about two rho masses. Beyond this region well-known ambiguities 
prevent definite results. Especially the onset of the continuous part
of the spectral functions at threshold can be reliably determined
and there are strong analogies to the results imposed on dispersion
theoretic approaches by the unitarity constraint.
On the other hand, the region near the physical nucleon-antinucleon
threshold $t=4M^2$ is most poorly defined, because through the boost
transformation it is an image of the nonrelativistic form factors
near $Q^2 \to \infty$. Both, $G_M^{(nr)}$ and $G_E^{(nr)}$, vanish
in this limit, so the possibility of finite results depends on 
the power of (diverging) factors $(1-t/4M^2)^{-1}$ introduced 
through the boost transformation. This does not seem a reliable way
for investigating this regime; it might prove more promising to try to
extract the timelike form factors above the $N$-$\bar N$ threshold
directly from soliton-antisoliton configurations.

\acknowledgments
The author thanks H.Walliser for many helpful discussions.

\newpage

\begin{figure}
\caption[]{
Proton magnetic and electric form factors $G_M^p$ and $G_E^p$
(divided by the standard dipole $G_D$) as obtained from the Skyrme
model with $e$=4.25, (soliton mass $M$=1648 MeV), and vector meson
coupling $\lambda$=0.75 in (\ref{eq:13}). The dashed lines show the
same form factors after subtraction of the square 
brackets in (\ref{eq:18}),(\ref{eq:19}). The data are from the
compilations of~\cite{Hoe76} (open circles),~\cite{Walker} (open
squares),~\cite{Sill} (full circles), and~\cite{Andi} (full triangles).}
\end{figure}

\begin{figure}
\caption[]{Neutron magnetic form factors $G_M^n$ (divided by the standard 
dipole $G_D$) and neutron electric form factor $G_E^n$ (full lines)
calculated as in fig.1. The data for $G_M^n$ are from~\cite{Lung} 
(full circles),~\cite{Bruins} (diamonds),
~\cite{Marko} (triangles), and~\cite{Anklin} (square). The open circles
represent older data from
~\cite{akerlof,albrecht,rock,stein,hughes,dunning}.
The data points for $G_E^n$ are from~\cite{Platchkov} for the Paris
potential (full circles) and from~\cite{Galster} 
for Feshbach-Lomon wave functions (open circles). 
The dash-dotted line for $G_E^n$ is the
Galster fit~\cite{Galster}, the dotted lines are fits 
from~\cite{Platchkov} deduced for three different potentials.}
\end{figure}

\begin{figure}
\caption[]{The continuous parts $\Gamma_{\rm cont}$ of the isoscalar
electric (full line) and magnetic (dashed line) spectral functions
for the skyrmion with $e$=4.25. In the upper part they are shown in
the soliton rest frame, in the lower part they are boosted to the 
Breit frame ($M$=1648 MeV).}
\end{figure}

\begin{figure}
\caption[]{Same as fig.3 for the isovector spectral functions.}
\end{figure}

\begin{figure}
\caption[]{The continuous part of the isovector electric spectral
function after inclusion of the $\varrho$-resonance.
The dotted line is the imaginary part of the $\varrho$-propagator
(with width 150 MeV);
the dashed line shows the continuous pionic part $\Gamma_{\rm cont}$ from 
(\ref{eq:65}) and fig.4; 
the dash-dotted line shows $\Gamma_E^{(T=1)}(t')$ 
as resulting from (\ref{eq:79}) (i.e. for $\lambda$=1).
The full line is the total result for $\lambda$=0.75.}
\end{figure}

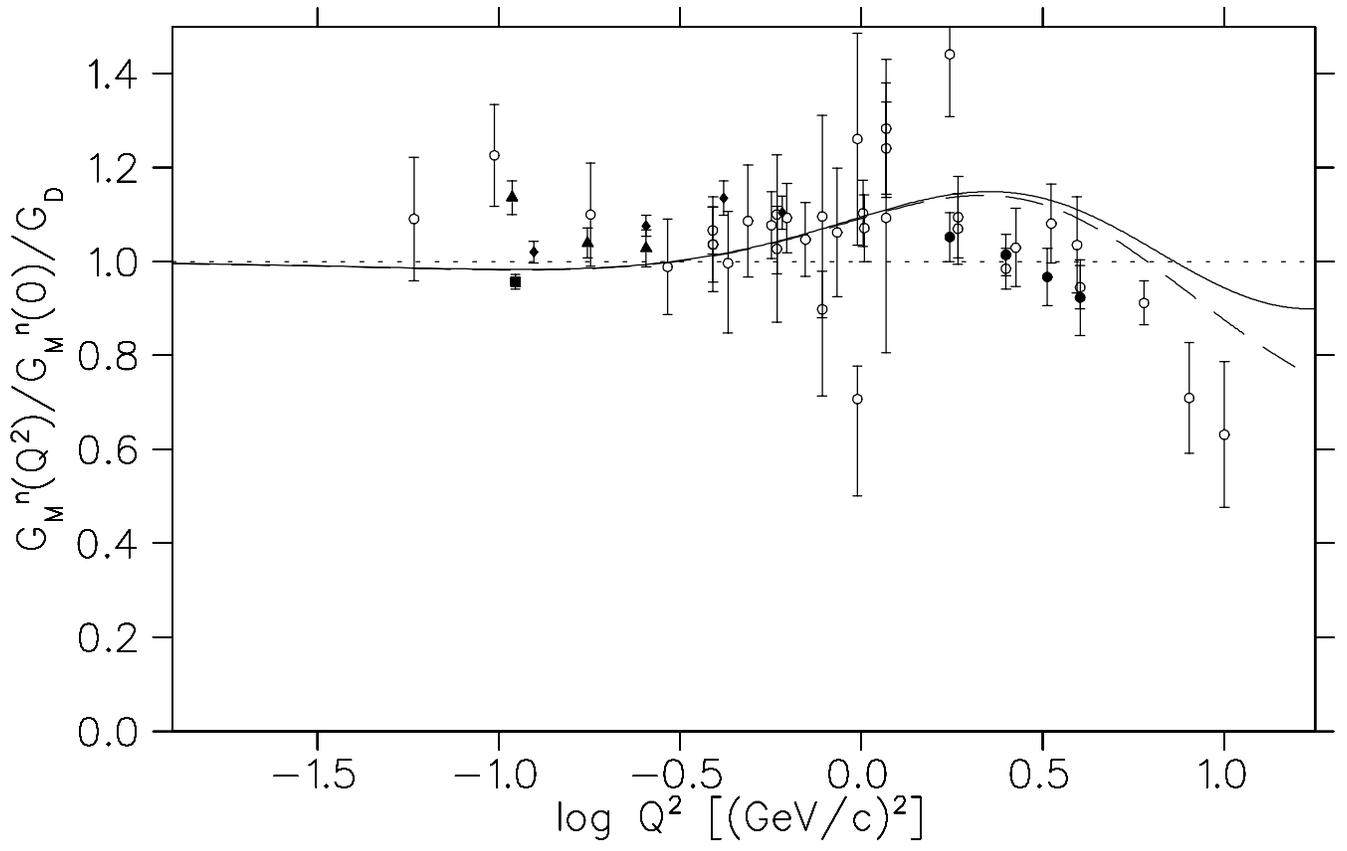
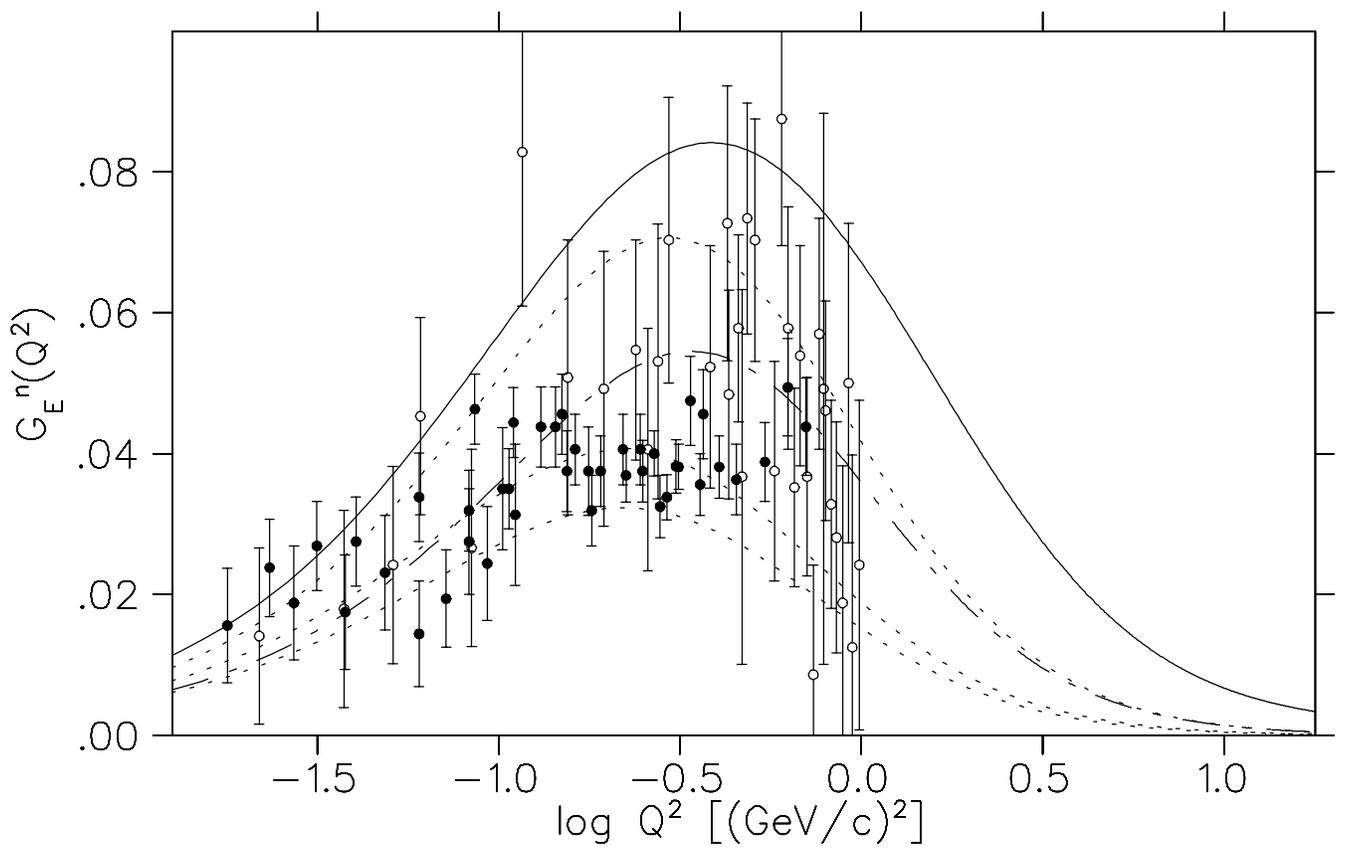
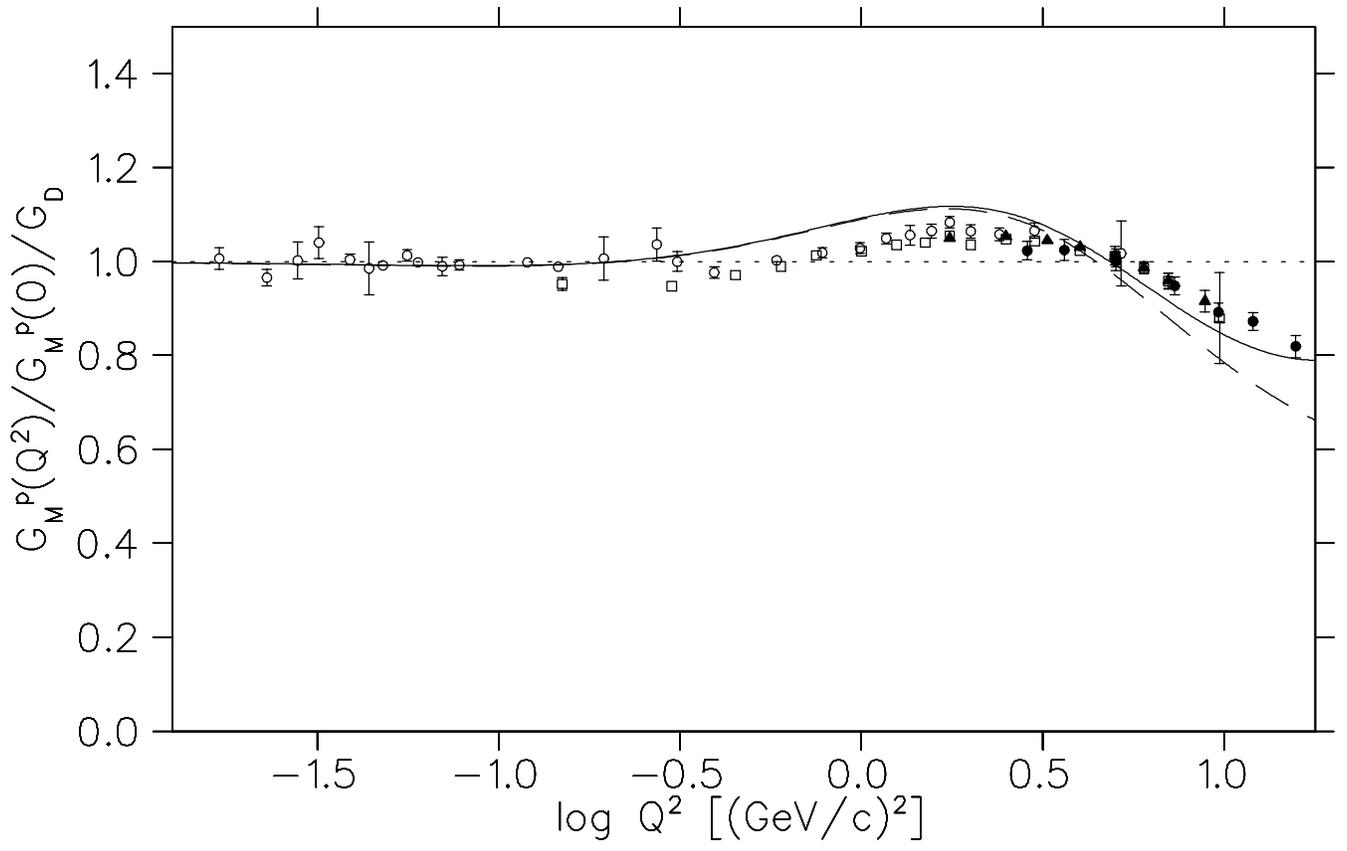
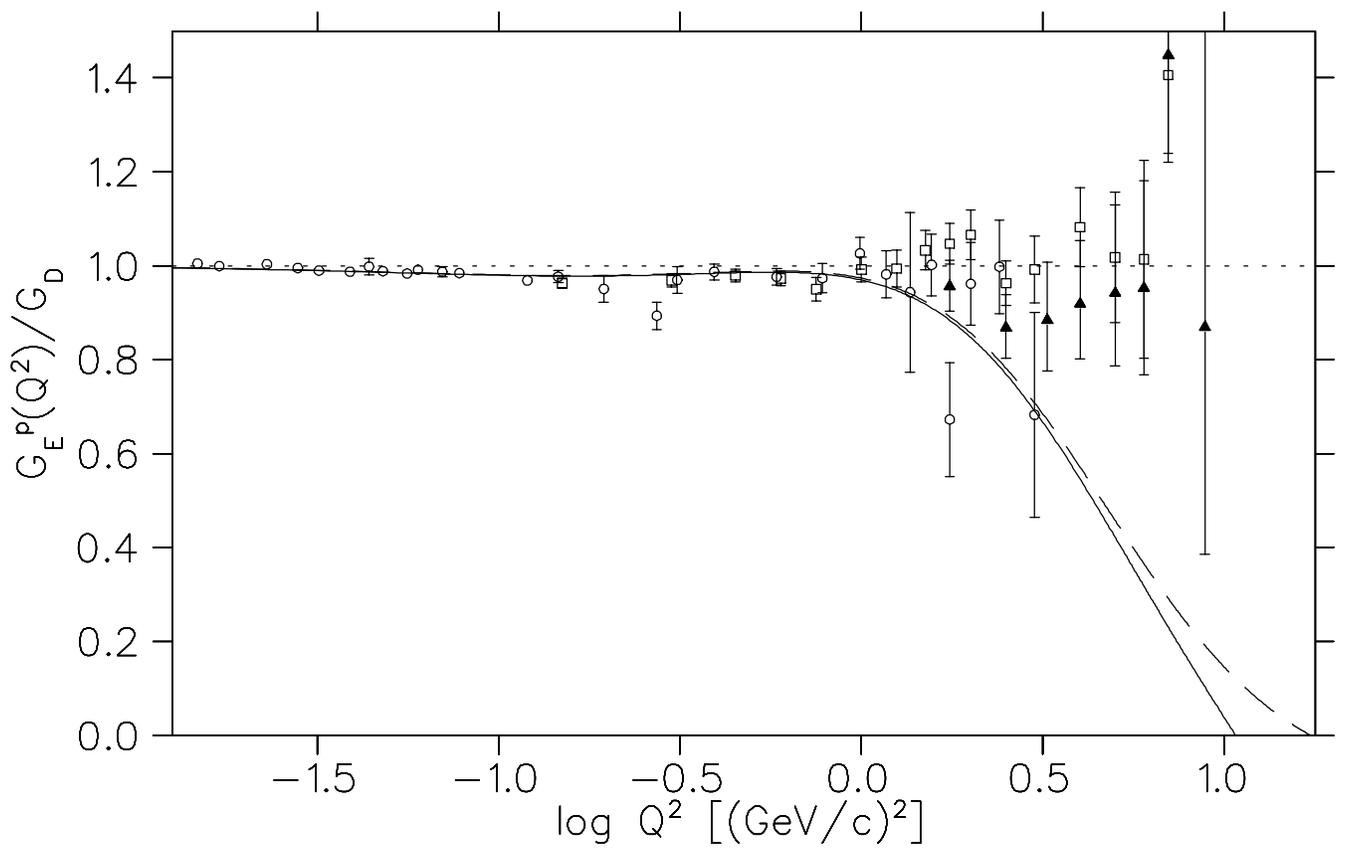
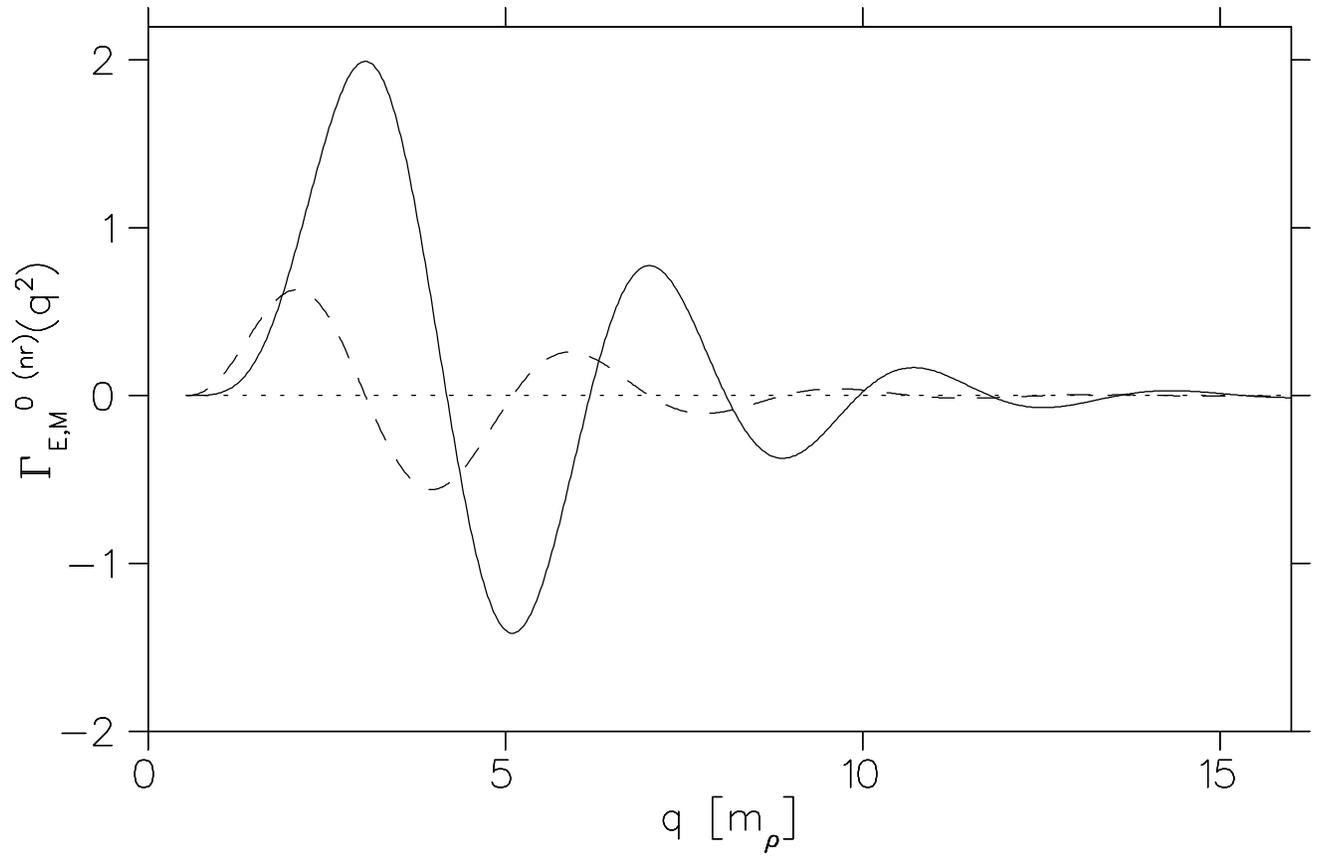
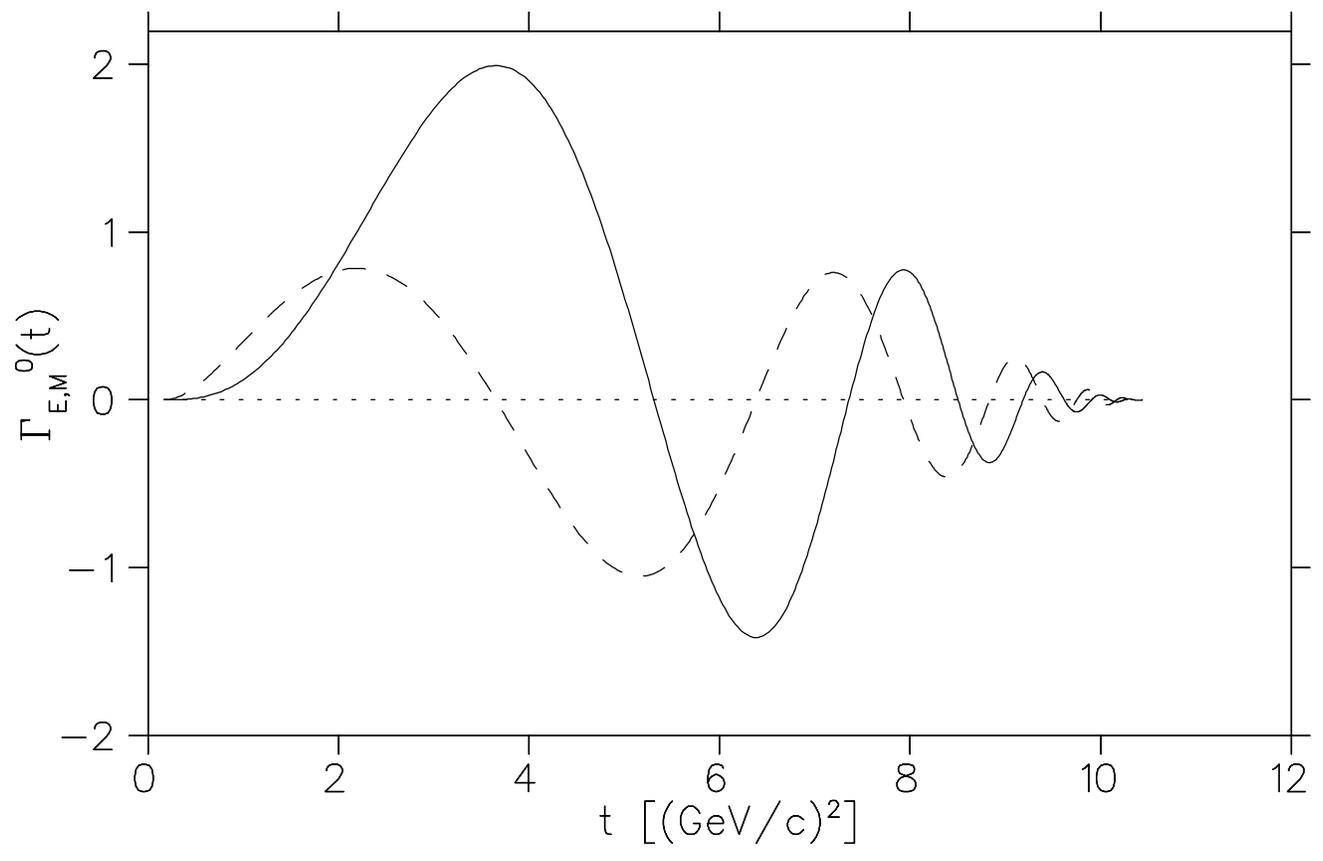
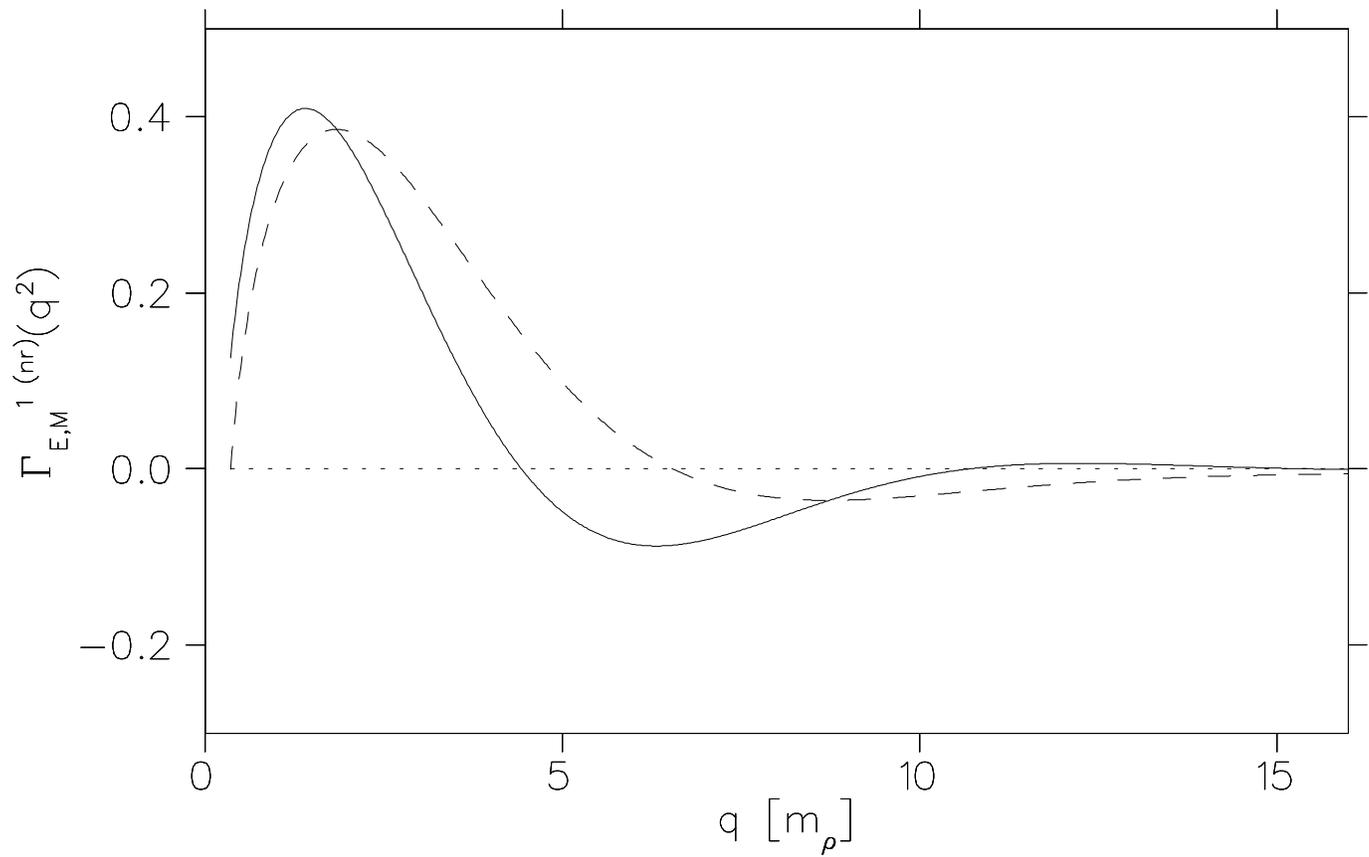
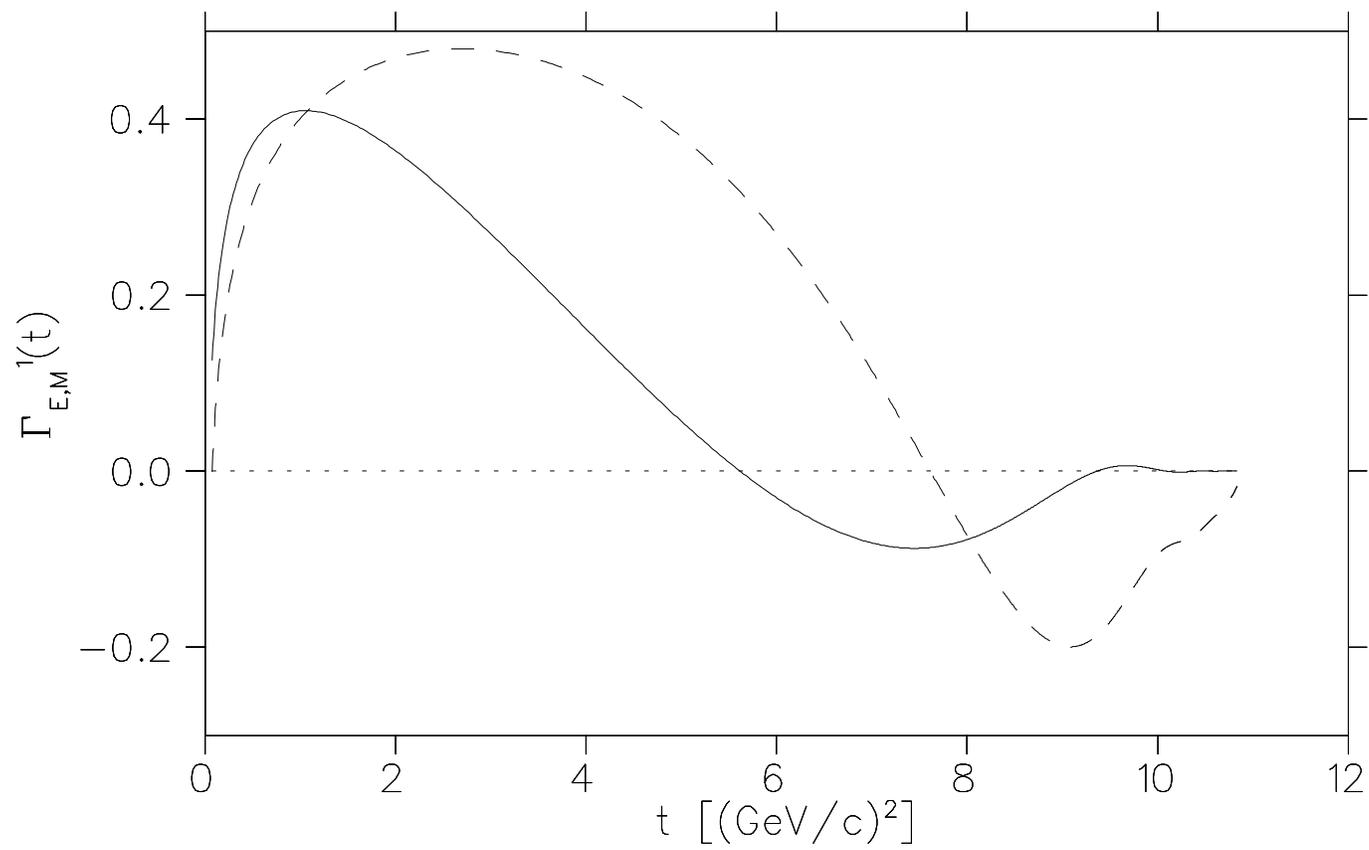
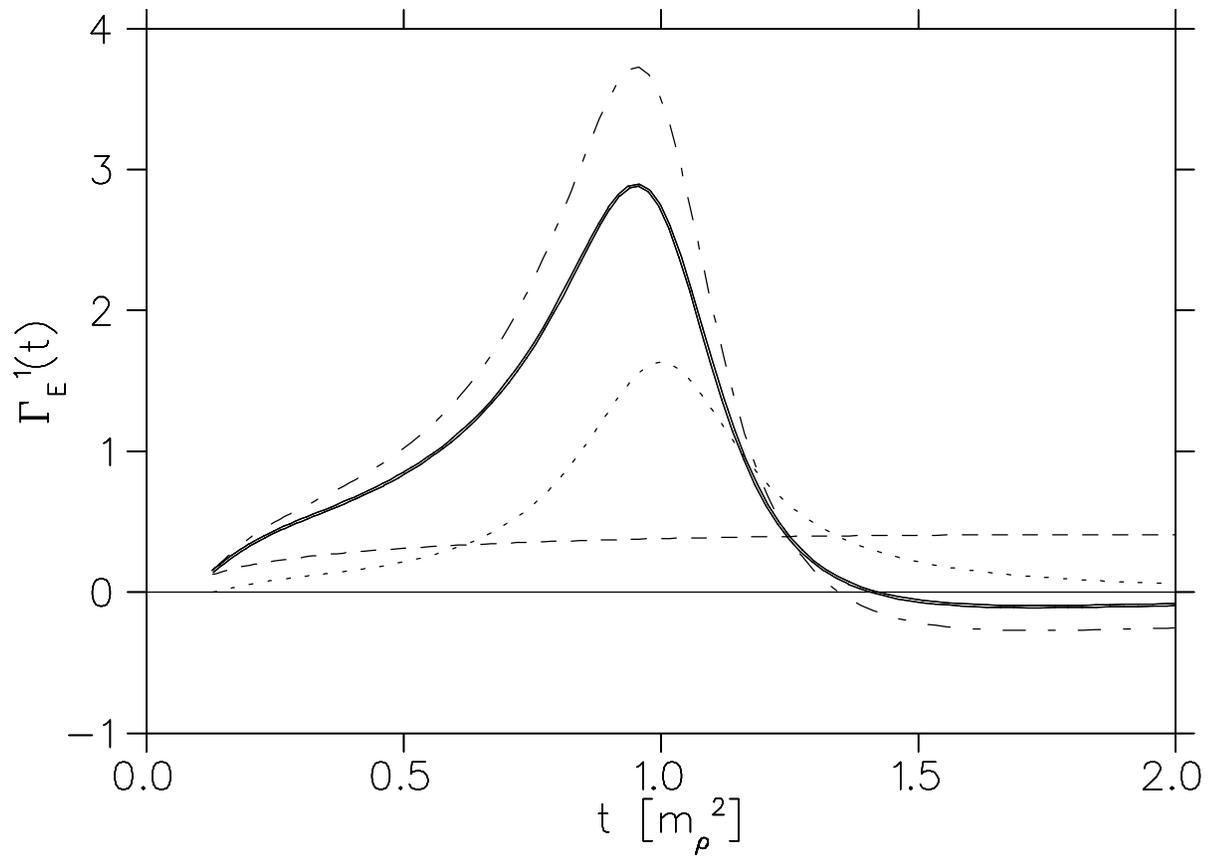
\end{document}